\input phyzzx

\pubnum={YITP 97-46}
\date={September 1997}

\titlepage
\title{ Geodesic flows for the Neumann-Rosochatius systems }

\author{
Reijiro Kubo$\dagger$,
Waichi Ogura$\ddagger$,
Takesi Saito$\S$
and Yukinori Yasui$\Vert$ 
}
\address{$\dagger$Yukawa Institute for Theoretical Physics \nobreak 
Kyoto University, Kyoto 606, Japan \break
$\ddagger$Department of Physics, Osaka University, Toyonaka, Osaka 565, Japan\break
$\S$Department of Physics, Kwansei Gakuin University, 
       Nishinomiya 662, Japan\break
$\Vert$Department of Physics, Osaka City University, Sumiyoshiku, Osaka 558,
Japan
}




\abstract{  
The Relationship between the Neumann system and the Jacobi system
 in arbitrary dimensions is elucidated from the point of view of 
constrained Hamiltonian systems. Dirac brackets for canonical 
variables of both systems are derived from the constrained 
Hamiltonians. The geodesic equations corresponding to the 
Rosochatius system are studied as an application of our method.
As a consequence a new class of nonlinear integrable equations 
is derived along with their conserved quantities.
 
}
\endpage
%
%
\leftline{\fourteenbf 1. Introduction}

To deal with Hamiltonian integrable systems, we usually rely on some basic
procedures such as determination of action-angle variables, separation
of variables, and integration of equations of motion. In addition,
it is often very instructive to study relations which exist among  
some superficially different mechanical systems. 

For instance, it is known that there is a close relationship between 
the Neumann system [1] and the Jacobi system [2]. The Neumann system 
describes the motion of a point particle on the $N-1$ sphere,
$S^{N-1}$, under the influence of a quadratic potential in 
$N$-dimensional space. Although the Neumann system has
been known for more than 100 years as an integrable nonlinear 
mechanical system, it is still under active study [3-7]. 
On the other hand, the Jacobi system describes
the geodesic flow on an ellipsoid. At first sight these two mechanical
systems are different. However, Kn{\" o}rrer [7] found the mapping
from the Jacobi system onto the Neumann system, showing that the two
systems are essentially equivalent. 

A more complicated problem was recently
studied by Adler and van Moerbeke [8]. They found that there are
rational maps transforming the Kowalewski top and the 
H${\acute e}$non-Heiles system into Manakov's geodesic flow on $SO$(4).   

In the present work we deal with some classical integrable systems
such as the Jacobi, Neumann, and Rosochatius systems laying special
emphasis on the relationship which exists among them. In sections 2
and 3, we deal with the Jacobi and the Neumann systems
in the framework of the constrained Hamilton formalism with a set of
second-class constraints. We calculate the Dirac brackets for canonical
variables in the Jacobi and the Neumann systems. We then transform 
variables so as to reduce second-class constraints to first-class
ones. As a consequence, both systems acquire gauge freedoms. 

The relationship between the Neumann system and the Jacobi system is
clarified by making use of the Gauss map introduced by Kn{\" o}rrer[7]. 
As an application of our method, we then study in section 4 the
geodesic equations corresponding to the Rosochatius system [9,10], 
which is also known as an integrable nonlinear system. 
Hamilton's equations for the $N$-dimensional Rosochatius system can 
be obtained from the $2N$-dimensional Neumann system. By the same
token, we found equations corresponding to the geodesic equations for the 
Rosochatius system from those of the $2N$-dimensional Jacobi
system. We show that the equations obtained are integrable by finding
integrals of motion explicitly. 
The last section is devoted to summary and concluding remarks.
\vskip 1cm


\leftline{\fourteenbf 2. The Neumann system }

In this section we deal with the famous Neumann system [1] in arbitrary
dimensions in the framework of the constrained Hamiltonian formalism 
with second-class constraints. Recently,various aspects of the Neumann 
system are investigated by many authors [3-7], because the Neumann
system is interesting as one of the simplest integrable non-linear 
systems and it is closely connected to other non-linear mechanical 
systems, such as those of the KdV, the Kowalewski top etc.. Our aim 
here is to clarify the Poisson structure of the Neumann system.

Let $(x_1,...,x_N,v_1,...,v_N)$ be canonical coordinates, so that 
the Poisson brackets are given by $\{ x_i,~v_j\}_P = \delta_{ij},
\{ x_i,~x_j\}_P = \{ v_i,~v_j\}_P = 0$,
 and the
symplectic form $\omega$ is written as $\omega = \sum_{k=1}^N
dx_k\wedge dv_k $. The Hamiltonian for the Neumann system is
written as
$$
{\cal H} = {\cal H}_N +\xi~\Omega_1 + \eta~\Omega_2 ,
\eqno(1)
$$
where $\xi$ and $\eta$ are Lagrange multipliers for the second-class
constraints
$$
\eqalign{
\Omega_1 &={1\over 2}\Bigl(\sum_{k=1}^N x_k^2 - r^2 \Bigr), \cr
\Omega_2 &= \sum_{k=1}^N x_kv_k \cr
}
\eqno(2)
$$
 and the classical Hamiltonian ${\cal H}_N$ for the Neumann system 
is given by
$$
{\cal H}_N = {1\over 2}\sum_{k=1}^N(v_k^2 + a_k x_k^2)
\eqno(3)
$$ 
with $r(\ne 0)$ and
$0<a_1<a_2<\cdots <a_N$ being real constant parameters. 

The Hamilton equations derived from (1) are
$$
\eqalignno{
&{\dot x}_k - v_k - \eta x_k = 0,&(4)\cr
&{\dot v}_k + a_kx_k +\xi x_k + \eta v_k = 0,&(5)\cr
&\sum_{k=1}^N x_k^2 - r^2 = 0,&(6)\cr
&\sum_{k=1}^N x_kv_k = 0.&(7)\cr
}
$$
It follows from (4) and (6) that
$$ 
\sum_{k=1}^N x_k{\dot x}_k = \sum_{k=1}^N x_kv_k + \eta~r^2 = 0,
\eqno(8)
$$
so that 
$$
\eta = 0.
\eqno(9)
$$
Also we have from (4),(5) and (7) 
$$
\sum_{k=1}^N(x_k{\dot v}_k + v_k{\dot x}_k)
= \sum_{k=1}^N(v_k^2 - a_kx_k^2) -\xi~r^2 = 0.
\eqno(10)
$$
It follows from (4),(5) and (9) that
$$
{\ddot x}_k = {\dot v}_k = - a_kx_k - \xi x_k.
\eqno(11)
$$
Combining (10) with (11) one obtains 
$$
{\ddot x}_k = - a_kx_k - {x_k\over r^2}\sum_{j=1}^N({\dot x}_j^2 - a_jx_j^2).
\eqno(12)
$$

The Dirac brackets [12,13] evaluated from (1) are given by 
$$
\eqalignno{
\{ x_i,~x_j\}_D &= 0,&(13)\cr
\{ x_i,~v_j\}_D &= \delta_{ij} - {x_ix_j\over x^2},&(14)\cr
\{ v_i,~v_j\}_D &= -{(x_iv_j -x_jv_i)\over x^2}, &(15)\cr
}
$$
where $x^2 \equiv \sum_{k=1}^N x_k^2 $.

Let us now introduce new variables $y_i,~(i=1,...N),$ defined by
$$
v_i = y_i - {x_i\over x^2}(x\cdot y),
\eqno(16)
$$
where $x\cdot y \equiv \sum_{k=1}^Nx_ky_k$.
By this change of variables, one obtains the identity
$$
x\cdot v \equiv \sum_{k=1}^Nx_kv_k = 0.
\eqno(17)
$$ 
Therefore, the term proportional to ${\eta}$ in the Hamiltonian (1)
vanishes identically.
We see that (7) also turns out to be an identity. It should be noted
here that the change of variables gives rise to a qualitative change
of the class of constraints. Namely, although $\Omega_1$ and
$\Omega_2$ were originally second-class constraints, after the change of
variables $\Omega_2$ vanishes identically and $\Omega_1$ turns out 
to be a first-class constraint with
the nonvanishing Lagrange-multiplier $\xi$. 
As a consequence,
the system acquires gauge freedom and $\xi$ becomes the gauge parameter. 

In order to see the gauge freedom explicitly, let us 
introduce the Lagrangian ${\cal L}$ for the Neumann system by 
the Legendre transform
$$
{\cal L} = \sum_{k=1}^N{\dot x}_kv_k - {\cal H}.
\eqno(18)
$$ 
Setting $\Omega_2 = 0$ and substituting (16) into (18),
 we obtain
$$
{\cal L}_{(\Omega_2=0)} = \sum_{k=1}^N{\dot x}_k(y_k + \lambda x_k)
-{1\over 2 x^2}\sum_{k>l}^NJ_{kl}^2- {1\over 2}\sum_{k=1}^N a_kx_k^2,
\eqno(19)
$$
where $\lambda = \zeta - ((x\cdot y)/x^2),~~{\dot\zeta}= \xi$ ,
 $J_{kl}= x_ky_l - x_ly_k$, and total derivative terms with respect to 
time have been deleted from (19).

The Euler equations obtained from (19) are
$$
\eqalignno{
{\dot x}_k &= y_k - {(x\cdot y)\over x^2}x_k, &(20)\cr
&\sum_{k=1}^N x_k{\dot x}_k = 0, &(21)\cr
{\dot y}_k &= -{\dot\lambda}x_k - a_kx_k +{1\over x^2}(x\cdot y)y_k .
&(22)\cr
}
$$
It follows from (20) and (22) that
$$
{\ddot x}_k = - a_kx_k - {x_k\over x^2}\sum_{l=1}^N({\dot x}_l^2 -
a_lx_l^2),
\eqno(23)
$$
where we have used the identity
$$
\sum_{k=1}^N{\dot x}_k^2 = y^2 - {(x\cdot y)^2 \over x^2}.
\eqno(24)
$$
Eq.(21) implies that
$$
x^2 = \sum_{k=1}^N x_k^2 = {\rm const..}
\eqno(25)
$$

On the other hand, changing variables from $v_k$ to $y_k$ by (16),
we have from (13),(14) and (15) that 
$$
\eqalignno{
\{ x_i,~x_j\}_P &= 0,&(26)\cr
\{ x_i,~y_j\}_P &= \delta_{ij} ,&(27)\cr
\{ y_i,~y_j\}_P &= -{(x_iy_j -x_jy_i)\over x^2}, &(28)\cr
}
$$
It should be noted that the Dirac brackets (13),(14),and (15) turn out
to be the simple Poisson brackets here.

Since $J_{kl}$ is invariant under the gauge transformation,
$$
(x_k,~y_k) \Longrightarrow (x_k,~y_k + \lambda x_k),
\eqno(29)
$$
where $\lambda$ is any function of $x$ and $y$, the Hamiltonian
$$
{\cal H}_N = {1\over 2x^2}\sum_{k>l}^NJ_{kl}^2 + {1\over 2}
\sum_{k=1}^N a_kx_k^2
\eqno(30)
$$
is also invariant under the gauge transformation (29), and so does
the equation of motion (23). Although the parameter $\lambda$ in (19) 
plays a role of a Lagrange multiplier in dealing with the
 constraint $\sum_{k=1}^Nx_k{\dot x}_k = 0$ imposed on the system, 
it can be regarded, on the other hand, as a gauge parameter. 
One notices that the change
of variables from $v_k$ to $y_k,~(k=1,...,N),$ gives rise to a
transmutation of constraints from the second-class to the first-class
ones, so that a gauge freedom appears through the 
gauge parameter $\lambda$ in the system. 


The Neumann system is Liouville integrable, because there exist
$N-1$ independent quantities [10], which are constants of motion
and which are in involution. In fact define the gauge-invariant
quantities[11] 
$$
F_k = x_k^2 + {1\over x^2}\sum_{l\ne k}^N{J_{kl}^2\over a_k-a_l}.
\eqno(31)
$$
Then, we can easily show that the $F_k~'s$ are conserved quantities,
 ${\dot F}_k =0$, and that they are in involution 
$$
\{ F_k,~F_l \}_P = 0,~~~(k,l = 1,...,N).
\eqno(32)
$$
$N-1$ of them are independent, because
$$
\sum_{k=1}^N F_k = x^2 = {\rm const}.. 
\eqno(33)
$$
The Hamiltonian for the Neumann system is expressed in terms of the
$F_k~'s$ as
$$
{\cal H}_N = {1\over 2}\sum_{k=1}^N a_kF_k.
\eqno(34)
$$
\vskip 1cm

\leftline{\fourteenbf 3. The Jacobi system}

The Neumann system is closely related to the mechanical system of
the geodesic motion of a particle on an ellipsoid, which was already  
shown to be integrable by Jacobi in 1838 [2]. The  Lagrangian of the 
geodesic motion is written as ${\cal L}= \sum_{k=1}^N q_k^{\prime}p_k
- {\cal H}$, where the Hamiltonian ${\cal H}$ of the geodesic motion
is given by
$$
{\cal H} = {\cal H}_J + { \xi\over 2}
\Bigl(\sum_{k=1}^N{q_k^2
\over b_k} - 1\Bigr) + \eta\sum_{k=1}^N{q_kp_k\over b_k} 
\eqno(35)
$$
with ${\cal H}_J = {1\over 2}p^2 = {1\over 2}\sum_{k=1}^N p_k^2$ . 
The prime ``~$^{\prime}$~'' denotes the differentiation with respect to
the parameter $s$, e.g. $q_k^{\prime}= dq_k/ds$.
The Hamilton equations derived from (35) are
$$
\eqalignno{
& q_k^{\prime} - p_k - \eta{q_k\over b_k} = 0,&(36)\cr
& p_k^{\prime} + \xi{q_k\over b_k} + \eta{p_k\over b_k} = 0,
&(37)\cr
&\sum_{k=1}^N {q_k^2\over b_k}-1  = 0,&(38)\cr
&\sum_{k=1}^N {q_kp_k\over b_k} = 0.&(39)\cr
}
$$
It follows from (36),(38) and (39) that
$$
\sum_{k=1}^N {q_k q_k^{\prime}\over b_k} = \sum_{k=1}^N {q_kp_k\over b_k}
 + \eta\sum_{k=1}^N{q_k^2\over b_k^2}= 0,
\eqno(40)
$$
so that 
$$
\eta = 0.
\eqno(41)
$$
Also we have from (36),(37) and (39) 
$$
\sum_{k=1}^N{q_k p_k^{\prime} + p_k q_k^{\prime} \over b_k}
= \sum_{k=1}^N{p_k^2\over b_k} -\xi\sum_{k=1}^N{q_k^2\over b_k^2} = 0.
\eqno(42)
$$
Thus we have
$$
q_k^{\prime\prime} =  p_k^{\prime} =  -\xi{q_k\over b_k} 
= - {1\over R^2}\biggl(\sum_{l=1}^N{p_l^2\over b_l} 
\biggr)~
{q_k\over b_k},
\eqno(43)
$$
where 
$$
R^2\equiv \sum_{l=1}^N(q_l^2/ b_l^2).
\eqno(44)
$$

The Dirac brackets are found to be
$$
\eqalignno{
\{ q_i,~q_j\}_D &= 0,&(45) \cr
\{ q_i,~p_j\}_D &= \delta_{ij} - {1\over R^2}{q_iq_j\over
b_ib_j},&(46) \cr
\{ p_i,~p_j\}_D &= -{1\over R^2}{(q_ip_j -q_jp_i)\over b_ib_j}. &(47)\cr
}
$$

In order to see the connection between the Neumann and the Jacobi
systems, let us introduce new variables $x_i,(i=1,...,N)$ and the time
variable $t$ by [7] 
$$
\eqalignno{
x_i &= {r\over R}{q_i\over b_i},&(48) \cr
{ds\over dt} &= \kappa R^2.&(49) \cr
}
$$
Here $\kappa^2 = (R^2\sum_{l=1}^N{p_l^2/b_l})^{-1}$ can be
shown to be a constant. Therefore, we have the relation 
$$
{1\over R^2}\biggl(\sum_{l=1}^N{p_l^2\over b_l} 
\biggr)~{\dot s}~^2 = 1,
\eqno(50)
$$
where ${\dot s} = ds/dt$. The nonlinear mapping from $q$ to $x$ by (48)
is a kind of the Gauss mapping, which maps the ellipsoid onto the sphere with
the radius $r$.  

Differentiating both sides of (48) with respect to $t$ twice, we
obtain
$$
\eqalign{
{\ddot x}_k &= - {r\over R^3}\biggl(\sum_{l=1}^N{p_l^2\over b_l} 
\biggr){\dot s}^2~{q_k\over b_k^2} + {r\over R}\biggl({\ddot s}-
{2R^{\prime}\over R}{\dot s}^2\biggr){q_k^{\prime}\over b_k} \cr
&- {r\over R^2}\biggl(R^{\prime\prime}{\dot s}^2 -{2(R^{\prime})^2
\over R}{\dot s}^2 + R^{\prime}{\ddot s}\biggr){q_k\over b_k}.\cr
}
\eqno(51)
$$
Also  differentiating both sides of
(49) with respect to $t$
one obtains
$$
 {\ddot s} = {2R^{\prime}\over R}{\dot s}~^2.
\eqno(52)
$$ 
Substitution of (49), (50) and (52) into (51) gives
$$
\eqalign{
{\ddot x}_k &= - {x_k\over b_k} - RR^{\prime\prime}
\biggl( \sum_{l=1}^N{p_l^2\over b_l}\biggr)^{-1}x_k \cr
 &= - {x_k\over b_k} -{x_k\over r^2}\sum_{l=1}^N ({\dot x}_l^2 
- {x_l^2\over b_l}).\cr
}
\eqno(53)
$$
Putting $1/ b_k = a_k$ and $r^2=x^2$ we finally arrive at Neumann's
equations again:
$$
{\ddot x}_k = - a_kx_k - {x_k\over x^2}\sum_{l=1}^N({\dot x}_l^2 -
a_lx_l^2).
\eqno(54)
$$ 

It should be noted that the first derivative of $x_k$ can be written
as
$$
{\dot x}_k = y_k - {(x\cdot y)\over x^2} x_k,   
\eqno(55)
$$
where 
$$
y_k = {r\over R}{p_k\over b_k}~{\dot s} + \mu x_k
    = {r\over Rb_k}~(p_k{\dot s} + \mu q_k)  
\eqno(56)
$$
with $\mu$ an arbitrary gauge parameter, which may be a function 
of $x$ and $y$. Eq.(55) is corresponding to
(20). Differentiating (56) with respect to $t$, one finds
$$
{\dot y}_k = -{1\over b_k}x_k + {(x\cdot y)\over x^2}y_k 
- 2\mu{(x\cdot y)\over x^2}x_k + \mu^2x_k
+ {\dot\mu}x_k 
\eqno(57)
$$
This equation agrees with (22), if one fixes the gauge in (22)
as follows:
$$
{\dot\lambda}= {y^2\over x^2}- 2\mu{(x\cdot y)\over x^2}+\mu^2 + {\dot\mu}
\eqno(58)
$$
with $a_k = 1/b_k$ .
Needless to say that Neumann's equations also follow from (55) and (57).    

It is to be noted that the Jacobi system is also Liouville integrable, 
because there exist $N$ gauge-invariant quantities $G_k,~(k=1,...,N),$ 
corresponding to $F_k$ in (31) [11],
$$
G_k = p_k^2 + \sum_{l\ne k}^N{K_{kl}^2\over b_k-b_l}
\eqno(59)
$$
with
$$
K_{kl} = q_kp_l - q_lp_k,
\eqno(60)
$$
which are also shown to be constants of motion, ${G_k}^{\prime}= 0$, and
$$
\{ G_k,~G_l\}_P = 0.
\eqno(61)
$$
$N-1$ of them are independent, because
$$
\sum_{k=1}^N{G_k\over b_k} = \biggl( 1 - \sum_{k=1}^N{q_k^2\over b_k}
\biggr)\sum_{k=1}^N{p_k^2\over b_k}- \biggl( \sum_{k=1}^N{q_kp_k\over b_k}
\biggr)^2= 0 .
\eqno(62)
$$

Hamiltonian for the Jacobi system can be written in terms of $G_k$
as follows:
$$
{\cal H}_J = {1\over 2}\sum_{k=1}^N G_k = {1\over 2} p^2.
\eqno(63)
$$

One finds from (31), (48), (56) and (59) the relation
$$
F_k = {r^2{\dot s}^2\over R^4b_k^2}(p_k^2 - G_k)
= -  {r^2 \kappa^2\over b_k^2}\sum_{l\ne k}^N{K_{kl}^2\over b_k-b_l},
\eqno(64)
$$
from which one obtains useful formulas such as
$$
\eqalignno{
\sum_{k=1}^N b_kF_k &= \sum_{k=1}^N{F_k\over a_k}= {r^2\over
R^2},&(65)\cr
\sum_{k=1}^N b_k^2 F_k &=\sum_{k=1}^N {F_k\over a_k^2} = 0.&(66)\cr
}
$$
It also follows from (50) and (56) that
$$
\sum_{k=1}^N {y_k^2\over a_k} = r^2 \Bigl( 1 + {\mu^2\over R^2}\Bigr).
\eqno(67)
$$
It is interesting to note that this equation is transformed to 
$$
\sum_{k=1}^N {z_k^2\over a_k} = r^2 \Bigl( 1 + {(\mu+\nu)^2\over R^2}\Bigr)
\eqno(68)
$$
under the gauge transformation $y_k\Longrightarrow z_k=y_k+\nu x_k$.
In consequence, the phase space of the Jacobi system is mapped 
by (48) and (56) onto the sphere defined by $x^2=r^2$ 
and the manifold given by (67) in the phase space of the Neumann
system. If one fixes the gauge parameter $\mu$, the mapping is 
obviously bijective.

\vskip 1cm
\leftline{\fourteenbf 4. The Rosochatius system}

In this section we generalize the formalism developed in preceding
sections and consider the Rosochatius system [9,10] in the generalized
framework.
Let us start with rewriting (48) in the form
$$
x_i = {r\over R}f_i(q),
\eqno(71)
$$
where $f_i(q),~(i=1,...,N),$ are regular functions of the coordinates
$q_j,~(j=1,...,N)$ which correspond to those introduced in the Jacobi
system. The function $R = R(q)$ is assumed to be given in terms of
$f_i(q)$ as follows:
$$
R^2 = \sum_{i=1}^N f_i^2(q)
\eqno(72)
$$
so that 
$$
\sum_{i=1}^N x_i^2 = {r^2\over R^2}\sum_{i=1}^N f_i^2 = r^2 = {\rm
const.}
\eqno(73)
$$
showing that $x_i,~(i=1,...,N)$ are variables on the $N-1$ sphere
 $S^{N-1}$.

Differentiating $x_i$ with respect to $t$, one obtains
$$
{\dot x}_i = y_i - {(xy)\over x^2}x_i ,
\eqno(74)
$$
where
$$
y_i = {r\over R}{\dot s}~f_i^{\prime} + \mu x_i
\eqno(75)
$$
with $\mu$ the gauge parameter, ${\dot s}= ds/dt$ and $f_i^{\prime}
= df_i/ds$.

Differentiating both sides of (74) once again with respect to $t$,
one obtains
$$
{\ddot x}_i = {\dot y}_i - {x_i\over x^2}({\dot x}^2 + x{\dot y})
-{(xy)\over x^2}y_i + {(xy)^2\over x^4}x_i.
\eqno(76)
$$
As we noticed in the case of the Neumann system, ${\dot y}_i$ is written
in terms of $x_i$ and $y_i$ as follows:
$$
{\dot y}_i = A_ix_i + By_i,
\eqno(77)
$$
where
$$ 
\eqalign{
B& = {(xy)\over x^2}, \cr
A_i &= -a_i,~~~~~(i=1,...,N).\cr 
}
\eqno(78)
$$

If we take
$$
A_i = -a_i + {c_i\over x_i^4}
\eqno(79)
$$
in addition to $B - (xy)/ x^2 = 0$, then (76) turns out to be
$$
{\ddot x}_i = -a_ix_i + {c_i\over x_i^3} 
 - {x_i\over x^2}\Bigl({\dot x}^2 - \sum_{k=1}^Na_kx_k^2 
+ \sum_{k=1}^N {c_k\over x_k^2}\Bigr),
\eqno(80)
$$ 
which are the equations of motion of a particle on the sphere
$S^{N-1}$ under the influence of the potential
$$
U(x) = {1\over 2}\sum_{k=1}^N\Bigl( a_kx_k^2 + {c_k\over x_k^2}\Bigr).
\eqno(81)
$$
Eqs.(80) are known as the Rosochatius equations, and the
Hamiltonian system governed by the potential $U(x)$ given above is called
the Rosochatius system [9,10]. This system has been shown to be completely
integrable by Rosochatius. 

The geodesic equations (43) in the Jacobi system is rewritten in terms of the
$f_i$'s as
$$
{d^2f_i\over ds^2} + {1\over R^2}a_if_i\sum_{j=1}^N{1\over a_j}
\Bigl({df_j\over ds}\Bigr)^2 = 0.
\eqno(82)
$$
Our final task is to find the geodesic equations for the Rosochatius
system corresponding to (82) in the Jacobi system. In order to do this
we first consider the 2$N$-dimensional Neumann system given
by the Hamiltonian
$$
{\cal H} = {\cal H}_{2N} + {1\over 2}\xi\Bigl(\sum_{i=1}^{2N}z_i^2 - r^2 \Bigr)
+ \eta\sum_{j=1}^{2N}z_jw_j,
\eqno(83)
$$
where
$$
{\cal H}_{2N} = {1\over 2}\sum_{k=1}^N[w_k^2 + w_{k+N}^2 
+ a_k z_k^2 + a_{k+N}z_{k+N}^2 ]
\eqno(84)
$$
with $w_i = {\dot z}_i~~(i=1,...,2N)$ and $a_{k+N}= a_k$ [3].    

We introduce new variables $x_k$ and $\theta_k,~(k=1,...,N),$ by
$$
z_k = x_k{\rm cos}\theta_k,~~~~~z_{k+N} = x_k{\rm sin}\theta_k
\eqno(85)
$$
and rewrite ${\cal H}_{2N}$ in terms of the new variables as
$$
{\cal H}_{2N} = {1\over 2}\sum_{k=1}^N({\dot x}_k^2 
+ x_k^2{\dot\theta}_k^2 + a_k x_k^2 ).
\eqno(86)
$$  
We then restrict ourselves to the specific case in which
the angular momenta $z_kw_{k+N} - z_{k+N}w_k,~(k=1,...,N)$ are
the constants of motion, that is,
$$
z_kw_{k+N} - z_{k+N}w_k = x_k^2{\dot\theta}_k = \sqrt{c_k}~(k=1,...,N) 
\eqno(87)
$$
with $c_k,~(k=1,...,N)$ real constants.
Substituting (87) into (86), we obtain the Hamiltonian for the
Rosochatius system
$$
{\cal H}_R = {1\over 2}\sum_{k=1}^N({\dot x}_k^2 
+{c_k\over x_k^2} + a_k x_k^2 ).
\eqno(88)
$$

We next consider the $2N$-dimensional Jacobi system. The Hamiltonian
is given by (35) with $N$ replaced by $2N$ and $b_k = b_{k+N}= 1/a_k,
(k=1,...,N)$.
The geodesic equations are then written as
$$
{d^2g_i\over ds^2} + {1\over R^2}a_ig_i\sum_{j=1}^N{1\over a_j}
\Bigl(\Bigl({dg_j\over ds}\Bigr)^2 + \Bigl({dg_{j+N}\over ds}\Bigr)^2
\Bigr) = 0,
\eqno(89)
$$
where 
$$
z_j = {r\over R}g_j,~~~~~z_{j+N} = {r\over R}g_{j+N}.
\eqno(90)
$$
In view of (85) it is convenient to introduce the variables
$f_i,~(i=1,...,N)$ by
$$
g_i = f_i~{\rm cos}\theta_i,~~~g_{i+N} = f_i~{\rm sin}\theta_i.
\eqno(91)
$$
Substituting (91) into (89), we obtain
$$
\eqalignno{
&f_i^{\prime\prime}- f_i(\theta_i^{\prime})^2 + {1\over R^2}a_if_i
\sum_{j=1}^N{1\over a_j}[(f_j^{\prime})^2 +
f_j^2(\theta_j^{\prime})^2]=0, &(92) \cr
&2f_i^{\prime}\theta_i^{\prime} + f_i\theta_i^{\prime\prime}=0.&(93) \cr
}
$$
Integrating (93) we find that $f_i^2\theta_i^{\prime}$ are constants
of motion. We set 
$$
f_i^2\theta_i^{\prime} = {\sqrt{d_i}\over b_i^2}.
\eqno(94)
$$
and if $f_j = q_j/b_j$ as in the $N$-dimensional Jacobi system, 
we finally obtain
$$
{d^2q_i\over ds^2} - {d_i\over q_i^3} +{1\over R^2}a_iq_i
\sum_{j=1}^N{1\over b_j}\Bigl[\Bigl({dq_j\over ds}\Bigr)^2 
+ {d_j\over q_j^2}\Bigr] = 0
\eqno(95)
$$ 
with $R^2$ given by (44).

It is interesting to note that $N$ integrals of motion
$H_k,(k=1,...,N),$ for the Rosochatius system are given by
$$
H_k = F_k + F_{k+N}= z_k^2 + z_{k+N}^2 +{1\over z^2}\sum_{l\ne k}^{2N}
{J_{kl}^2\over a_k-a_l} + {1\over z^2}\sum_{l\ne k+N}^{2N}
{J_{k+N,l}^2\over a_{k+N}-a_l},
\eqno(96)
$$   
where $J_{kl}= z_kw_l - z_lw_k$. It should be noted that both $F_k$
and $F_{k+N}$ contain singular terms because the Hamiltonian (83) is 
degenerate, $a_k = a_{k+N},(k=1,...,N
)$.
However, such singular terms cancel in $F_k + F_{k+N}$ and we have
$$
H_k = x_k^2 +{1\over x^2}\sum_{l\ne k}^{N}
{H_{kl}^2\over a_k-a_l},
\eqno(97)
$$
with[9,10]
$$
H_{kl}^2 = (x_ky_l-x_ly_k)^2 + {c_kx_l^2\over x_k^2} + {c_lx_k^2\over x_l^2}.
\eqno(98)
$$

In a similar way we obtain $N$ integrals $I_k,(k=1,...,N)$ for the
equations (95) combining pairs of integrals for the $2N$-dimensional
degenerate Jacobi system
$$
I_k = G_k + G_{k+N} = \pi_k^2 + \pi_{k+N}^2 
+ \sum_{l\ne k}^{2N}{K_{kl}^2\over b_k-b_l}
+ \sum_{l\ne k+N}^{2N}{K_{k+N,l}^2\over b_{k+N}-b_l},
\eqno(99)
$$
where $K_{kl}= \zeta_k\pi_l - \zeta_l\pi_k$ with 
$$
\eqalign{
\zeta_k &= b_kg_k = b_kf_k{\rm cos}\theta_k = q_k{\rm cos}\theta_k, \cr
\zeta_{k+N} &= b_{k+N}g_{k+N} = q_k{\rm sin}\theta_k, \cr
\pi_k & = \zeta_k^{\prime} = (p_k{\rm cos}\theta_k
- q_k{\theta}_k^{\prime}{\rm sin}\theta_k),\cr
\pi_{k+N} & = \zeta_{k+N}^{\prime} = (p_k{\rm sin}\theta_k
+ q_k{\theta}_k^{\prime}{\rm cos}\theta_k).\cr
}
\eqno(100)
$$  
Substituting (100) into (99), we find
$$
I_k = p_k^2 + {d_k\over q_k^2} + \sum_{l\ne k}^{N}{I_{kl}^2\over b_k-b_l}
\eqno(101)
$$
with
$$
I_{kl}^2 = (q_kp_l-q_lp_k)^2 + {d_kq_l^2\over q_k^2} + {d_lq_k^2\over q_l^2}.
\eqno(102)
$$
Since it can be shown that 
$$
{dI_k\over ds} = 0,~~~(k=1,...,N),
\eqno(103)
$$
$I_k$ are indeed conserved quantities of the system governed by (95).     
\vskip 1cm
\leftline{\fourteenbf 5.  Summary and concluding remarks}

In the first part of this work we dealt with the Neumann
and the Jacobi systems in the classical framework of constrained 
Hamiltonian systems. We calculated Dirac brackets for canonical 
variables. We noticed that a transmutation from the second-class 
constraints to the first-class ones occurred by changing dynamical
variables appropriately. As a consequence, both systems acquired
gauge freedom in terms of residual gauge-parameters.  

We focused on the relationship 
between the Neumann and the Jacobi systems. The mapping from the 
phase space of the Jacobi system to that of the Neumann system was
executed by the Gauss map. Affine connections appearing in the
geodesic equations in the Jacobi system are given by
$$
\Gamma^i_{jk} = {1\over R^2}{q_i\over b_i}{1\over b_j}\delta_{jk}.
\eqno(104)
$$ 
However, these affine connections do not satisfy the identities which
genuine affine connections should satisfy, because the geodesic
equations are not all independent. 

In the second part of the work we considered the Rosochatius
system. We derived the Hamiltonian ${\cal H}_R$ for the Rosochatius
system from the 2$N$-dimensional Neumann Hamiltonian. The geodesic 
equations for the Rosochatius system should be obtained by making use
of the 2$N$-dimensional Jacobi equations. However, the final result given
by (95) is by no means the geodesic equation for $q_i$, because (95)
does not have the standard form of the geodesic equation, which is 
usually written as
$$
\eqalign{
{d^2f_k\over ds^2} + &\Gamma^k_{ij}{df_i\over ds}{df_j\over ds}= 0,\cr
&\Gamma^k_{ij} = {\phi_k\over \sum_{l=1}^N\phi_l^2}
\sum_{i,j=1}^N\phi_{ij}, \cr
}
\eqno(105)
$$
where $\phi$ is the regular function of $f_i,~(i=1,...,N)$, 
$\phi_i= \partial\phi/\partial f_i$ and $\phi_{ij}= \partial^2\phi/
\partial f_i\partial f_j$.  

However, we have 
$$
\eqalign{
&{d^2\theta_k\over ds^2}
+ {2\over q_k}{dq_k\over ds}{d\theta_k\over ds}= 0, \cr
&{d^2q_k\over ds^2} - q_k\Bigl({d\theta_k\over ds}\Bigr)^2 
+{1\over R^2}a_kq_k
\sum_{j=1}^N{1\over b_j}\Bigl[\Bigl({dq_j\over ds}\Bigr)^2 
+ q_j^2\Bigl({d\theta_j\over ds}\Bigr)^2\Bigr] = 0\cr
}
\eqno(106)
$$
from (93),(94) and (95). These equations are geodesic equations
in the 2$N$-dimensional space described by the coordinates
$(q_k,~\theta_k),~(k=1,...,N)$.

We found that (95) are integrable and $I_k,~(k=1,...,N)$ are 
conserved quantities of the system governed by (95). In other words, 
the set of equations given by (95) constitutes an integrable
system dual to the Rosochatius system.

There are some important problems on the Rosochatius system with its
dual system left unsolved, such as finding classical solutions  
in arbitrary dimensions
and the problem of quantizing the Jacobi-Neumann and the Rosochatius 
systems.  Some classical solutions to the Neumann
system have been found by several
authors [6]. Quantization of the Neumann system is 
discussed in some extent by Gurarie [14]. We would like to discuss 
extensively those remaining problems in a forthcoming paper. 
\eject

\centerline{\fourteenbf Acknowledgement} 

We would like to thank Professor K. Takasaki for giving us 
a series of lectures on the Jacobi-Neumann systems. We also
thank Professor R. Sasaki for critical reading of the manuscript
and for valuable comments and discussions.  
\vskip 1cm

\centerline{\fourteenbf References}
\item{[1]} Neumann C 1859 {\it J. Reine Angew. Math.} {\bf 56} 46-63
\item{[2]} Jacobi C G J~1884 {\it Vorlesungen {\" u}ber Dynamik}~~in~ {\it
           Gesammelte Werke~Supplementband} (Berlin)
\item{[3]} Moser J 1979 {\it Geometry of Quadrics and Spectral Theory}(Chern
Symposium)(Berkeley) 147-188;\hfill\break
Moser J 1978 {\it Prog. Math.}{\bf 8} 233-289; \hfill\break
Moser J 1981 Integral Hamiltonian Systems and Spectral Theory
{\it Pisa,Lezioni Fermiane}
\item{[4]} Avan J and Talon M 1991{\it Phys.Lett.}{\bf B268} 209-216;
\hfill\break
Avan J and Talon M 1990 {\it Int.J.Mod.Phys.} {\bf A5} 4477-4488;\hfill\break 
Avan J and Talon M 1991 {\it Nucl.Phys.}{\bf B352} 215-249
\hfill\break
Avan J, {\it Phys.Lett.}{\bf A156} 61-68\hfill\break
Babelon O and Talon M 1992 {\it Nucl. Phys.} {\bf B379} 321-339 
\item{[5]} Ragnisco O and Suris Y B 1996 On the $r$-Matrix Structure of
the Neumann System and its Discretizations, in {\it Algebraic Aspects
of Integrable Systems} ed Fokas A S and Gelfand I M 
{\it Prog.Nonlinear Diff.Eqs. and Their Applications} (Birkh{\" a}user)
\hfill\break
Ratiu T (1981) {\it Trans. Amer. Math. Soc.} {\bf 264} 321-329 
\item{[6]} Mumford D 1983 {\it Tata Lectures on Theta II}~(Birkhaeuser)
\hfill\break
Dubrovin B A 1981 {\it Russ. Math. Surv.} {\bf 36} 11-80\hfill\break
Semenov-Tian-Shansky M A 1994 {\it Integrable Systems II}~ in~{\it Dynamical
Systems VII}~~ed Arnol'd V I and Novikov S P
(Verlin: Springer-Verlag)\hfill\break
Veselov A P 1980 {\it Funct. Anal. Appl.} {\bf 14} 37-39\hfill\break
Devaney R L 1976 {\it Amer. J. Math.} {\bf 100}  631-642
\item{[7]} Kn{\" o}rrer H 1982 {\it J. Reine Angew. Math.} {\bf 334} 69-78.
\item{[8]} Adler M and van Moerbeke P 1988 {\it Commun. Math. Phys.} 
{\bf 113} 659-700.
\item{[9]} Rosochatius E 1877 {\it {\" U}ber die Bewegung eines Punktes},
(Inaugural Dissertation, Universit{\" a}t G{\" o}ttingen, Gebr. Unger)
(Berlin) 
\item{[10]} Ratiu T 1982
The Lie Algebraic Interpretation of the Complete
Integrability of the Rosochatius System  in {\it Mathematical Methods 
in Hydrodynamics and Integrability in Dynamical Systems,} {\it AIP 
Conference Procs.}{\bf 88} 109-115 (AIP, New York) \hfill\break
Gagnon L, Harnad J and  Winternitz P 1985 {\it J. Math. Phys.} 
{\bf 26} 1605-1612
\item{[11]} Uhlenbeck K K 1982 {\it Lecture Notes in Mathematics} {\bf
949} 146-158
\item{[12]} Dirac P A M 1950 {\it Can. J. Math.} {\bf 2}
129-148;\hfill\break 
Dirac P A M 1951 {\it
Can. J. Math.} {\bf 3} 1-23; \hfill\break
Dirac P A M 1958 {\it Proc. Roy. Soc.} {\bf A246} 326-338;\hfill\break
Dirac P A M 1967 {\it Lectures on Quantum Mechanics}\hfill\break
(New York: Yeshiva University, Academic Press)
\item{[13]} Henneaux M and Teitelboim C 1992 {\it Quantization of Gauge
Systems}~ (Princeton: Princeton University Press)\hfill\break
Sundermeyer K 1982 {\it Constrained Dynamics, Lecture Notes in Physics}
{\bf 169}\hfill\break(Berlin, Springer) \hfill\break
Hanson A, Regge T and Teitelboim C 1976 {\it Constrained
Hamiltonian Systems~~~ Accad. Naz. dei
Lincei} (Rome) \hfill\break
Kugo T 1989 {\it The Quantum Theory of Gauge Fields} I, II,~
( Tokyo: Baifukan), ( in Japanese ), and references therein.
\item{[14]} Gurarie D 1995{\it J. Math. Phys.} {\bf 36} 5355-5391
\bye